\documentclass[12pt]{article}
\usepackage[left=2cm,right=2cm,top=2cm,bottom=2cm]{geometry}
\usepackage{amsmath, amssymb,amsfonts}
\usepackage{graphicx,color}
\usepackage{indentfirst}
\usepackage{color}
\usepackage{setspace}
\usepackage{axodraw}

\date{}
\title{Quantum Computation and Non-Abelian Statistics in Chern-Simons-Higgs Theory }
\author{{J. C. Brozeguini$^{1, 2}$ and E. C. Marino$^{1}$}\\
{\small\it $^{1}$Instituto de F\'\i sica, Universidade Federal do Rio de Janeiro, Cx. P. 68528}\\ {\small\it Rio de Janeiro, 21941-972 RJ Brazil}\\
{\small\it $^{2}$Instituto Federal de Educa\c c\~ao, Ci\^encia e Tecnologia do Rio de Janeiro} \\ {\small\it  Nil\'opolis, 26530-060 RJ Brazil}
}

\begin{document}
\begin{singlespace}
\maketitle
\begin{abstract}
We naturally obtain the NOT and CNOT logic gates, which are key pieces of quantum computing algorithms, in the framework of the 
non-Abelian Chern-Simons-Higgs theory in two spatial dimensions. For that,
we consider the anyonic quantum vortex topological excitations occurring in this system and
 show that self-adjoint (Majorana-like)
combinations of these vortices and anti-vortices have in general non-Abelian statistics. The associated unitary
monodromy braiding matrices become the required logic gates in the special case when the vortex spin is $s=1/4$.
We explicitly construct the vortex field operators, show that they carry both magnetic flux and charge
and obtain their euclidean correlation functions by using the method of quantization of topological excitations, which is
based on the order-disorder duality. These correlators are in general multivalued, the number of sheets being determined by the 
vortex spin. This, by its turn, is proportional to the vacuum expectation value of the Higgs field and therefore can be tuned both by
 the free parameters of the Higgs potential and the temperature.

\end{abstract}
\end{singlespace}

\leftline{PACS: 11.15.Yc, 03.67.-a, 03.67.Pp }

\section{Introduction}

Quantum computation is a subject that is strongly attracting the interest of the physics community in recent times \cite{qc-nas1,qc-nas2}, mainly because of the
 vast potential it has for extremely high-performance calculations. The main reason for that derives
from the fact that instead of employing a binary system as the basic computing unit, it uses the infinitely many quantum states obtained by coherent
linear combinations of certain base states. Loss of quantum coherence through interaction with the
environment, however, is a fatal threat for a quantum computer, since it implies complete loss of information.
For this reason it is important to find systems, capable of performing the required operations of quantum computation, being at the same time
protected against the process of decoherence.

Systems presenting non-Abelian statistics can provide the required stability through the mechanism known as topological quantum computation \cite{qc-nas2}. For
these systems, exchanging particles in a many-particle state produces the entanglement thereof, thus making them robust against decoherence. An intense
search for many-particle systems presenting such peculiar behavior under the particle exchange -- or braiding --  operation
has then started.

In this work, we consider the non-Abelian Chern-Simons field in 2+1 D, minimally coupled to a Higgs field in the adjoint representation of the SU(2) group
(the CSH-theory).
The Higgs field potential is such that there are two phases according to whether the vacuum expectation value of this field vanishes or not.
We study in detail the quantum magnetic vortex excitations of this system in the ordered phase. In order to accomplish the full quantization
of such excitations, we apply to the CSH-theory the method of quantization of topological excitations, which is based on the concept of order-disorder duality
\cite{marswi,marino-topexc1,marinoap,marino-topexc2}.
We obtain, in particular, the explicit form of the creation operator of quantum excitations carrying both magnetic flux and charge, as well as their
Euclidean correlation functions. These electrically charged magnetic vortices may be boson, fermion or, more generally anyons.

We show that special self-adjoint combinations of vortices and anti-vortices possess non-Abelian statistics whenever the
vortices are anyonic. Furthermore, for a specific value of the anyon spin, namely $s=1/4$, we show that we can construct the NOT and CNOT logic gates,
required in quantum computation, from the corresponding monodromy matrices. Our results, based on a fully quantized approach of
topological excitations, therefore show that the CSH model with an SU(2) group is an excellent example of a system exhibiting the requisites needed for the operation of
a quantum computer.

Related results have been reported in the literature. For instance, non-Abelian statistics has been obtained for certain
quasi-particle excitations of the quantum Hall liquid \cite{qhl} and also for Ising anyons \cite{NAS,IA}. Models inspired
in non-Abelian anyons have been proposed in \cite{naa}.

\section{The Chern-Simons-Higgs Theory}
\subsection{The Theory}

Let us consider the SU(2) non-Abelian Chern-Simons theory to which we couple a Higgs field in the
adjoint representation:
\begin{eqnarray}
\label{CSH}
S_{CS}[A] = \frac{\kappa}{4\pi}\int d^{3}z\,\epsilon^{\mu\nu\rho}\left(A^{a}_{\mu}
\partial_{\nu}A^{a}_{\rho} +\frac{2}{3}\,\epsilon^{abc}A^{a}_{\mu}
A^{b}_{\nu}A^{c}_{\rho}\right) + \mbox{Tr}D_{\mu}\Phi D^{\mu}\Phi - V(\vert \Phi\vert, \eta)
\end{eqnarray}
where the Higgs self-interaction potential is given by
\cite{hong1990multivortex,jackiw1990self,navarro2009non}
\begin{eqnarray}
V = (4\lambda)^{2}\mbox{Tr}\Phi^{2}(\eta^{2} + \Phi^{2})^{2}.
\end{eqnarray}

The Euler-Lagrange equation is
\begin{eqnarray}
\label{EL}
\frac{\kappa}{2\pi}\,\epsilon^{\mu\nu\rho}D^{ac}_{\nu}A^{c}_{\rho} = J^{\mu a}
\end{eqnarray}
where $J^{\mu a} = -2[\Phi, D^{\mu}\Phi]^{a}$.

The theory presents ordered or disordered phases, according to whether $\eta^2<0$ or $\eta^2>0$, where, respectively, we have
$\langle \Phi \rangle \neq 0$ and $\langle \Phi \rangle = 0$.

\subsection{Charge and Magnetic Flux Carrying Operators}

We are now going to obtain the operators $\sigma$ and $\mu$, which create states carrying, respectively, charge and magnetic flux. These two
quantities are given respectively by
\begin{eqnarray}
Q = \int d^{2}x\,J^{0 a}n^{a} \quad \mbox{and} \quad \Phi_{M} = \int d^{2}x\,B^{a}n^{a}
\end{eqnarray}
where $n^{a}$ is an unit vector, subject to the action of the group
(for instance $n^a=\frac{\phi^a}{\vert \phi \vert}$ where $\phi^a \equiv \langle \Phi^a \rangle$) and
\begin{eqnarray}
J^{0a} = \frac{\kappa}{2\pi}\epsilon^{ij}D^{ac}_{i}A^{c}_{j}(x) \qquad
\mbox{and}\qquad B^{a}(x) =\frac{1}{2} \epsilon^{ij}\,F_{ij}^{a}
\end{eqnarray}

In order to construct the local operators $\sigma$ and $\mu$, we will follow the method for 
quantization of topological excitations that was developed with basis on the order-disorder duality \cite{marswi,marino-topexc1,marinoap,marino-topexc2}. According to this, the $\sigma$ and $\mu$ operators act, respectively, as
order and disorder operators and therefore satisfy the corresponding dual algebra. Then, correlation functions 
determined by this algebra
 are obtained by coupling certain special external fields to the dynamical fields of the system. In the case of the
$\mu$-operator these external fields are given by
$\bar{A}^{b}_{\mu}(z;x_{1},\ldots, y_{M}) = \bar{A}^{b}_{\mu}(z; x_{1},\ldots,x_{N}) - \bar{A}^{b}_{\mu}(z; y_{1},\ldots,y_{M})$, whereas, for the $\sigma$-operator, by $\bar{C}_{\mu}^{ d}(z; x_{1},\ldots,y_{M}) = \bar{C}_{\mu}^{ d}(z; x_{1},\ldots,x_{N}) -
\bar{C}_{\mu}^{ d}(z; y_{1},\ldots,y_{M})$
where
\begin{eqnarray}
\bar{A}^{\mu b}(z; x_{1},\ldots,x_{N}) = a\sum_{i = 1}^{N}\mbox{arg}(z - x_{i})n^{b}
\int_{S_{x_{i}}} d^{2}\xi^{\mu}\,\delta^{3}(z -\xi),
\label{campoexterno1}
\end{eqnarray}

and

\begin{eqnarray}
\bar{C}^{\mu d}(z; x_{1},\ldots,x_{N}) =
b\sum_{i = 1}^{N}\mbox{arg}(z - x_{i})n^{d}\int_{S_{x_{i}}}
d^{2}\xi^{\lambda}\,\epsilon^{\lambda\mu\nu}\partial_{\nu}\delta^{3}(z - \xi).
\label{campoexterno2}
\end{eqnarray}

In the above expressions $d^{2}\xi^{\mu} = \frac{1}{2} \epsilon^{\mu\alpha\beta}(d \xi_\alpha d \zeta_\beta-d \xi_\beta d \zeta_\alpha)$ is the covariantized vector surface integration element, perpendicular to the integration  surface
$S_{x_{i}}$. This consists of the complex plane, excluding the singularities at $x_i$ and along the cut
of the function $\mbox{arg}(z - x_{i})$.

It turns out that the mixed multicorrelation function is given by the vacuum functional in the 
presence of these external fields:
\begin{eqnarray}
\label{ff}
& &
\langle \sigma(x^{a}_{1})\mu_{R}(x^{b}_{1})\ldots \sigma(x^{a}_{N})\mu_{R}(x^{b}_{N})\mu^{\dagger}_{R}(y^{b}_{M})\sigma^{\dagger}(y^{a}_{M})
\ldots\mu^{\dagger}_{R}(y^{b}_{1})\sigma^{\dagger}(y^{a}_{1})\rangle =
\nonumber\\
& &
{\cal Z}^{-1}\int {\cal D} A^{a}_{\mu}{\cal D}\Phi^{b}{\cal D}\eta {\cal D}
\bar{\eta}\,\exp\Bigg\{
-\int d^{3}z\Bigg[
\frac{\kappa}{4\pi}\epsilon^{\mu\nu\rho}\Big[[A^{d}_{\mu} + \bar{A}^{d}_{\mu}(x^{b}_{1},
\ldots, y^{b}_{M}) + \bar{C}^{d}_{\mu}(x^{a}_{1}, \ldots, y^{a}_{M})]
\partial_{\nu}[
\nonumber\\
& &
A^{d}_{\rho} + \bar{A}^{d}_{\rho}(x^{b}_{1}, \ldots, y^{b}_{M}) +
\bar{C}^{d}_{\rho}(x^{a}_{1}, \ldots, y^{a}_{M})]
+\frac{2}{3}\,\epsilon^{def}[A^{d}_{\mu} + \bar{A}^{d}_{\mu}(x^{b}_{1}, \ldots, y^{b}_{M}) +
\bar{C}^{d}_{\mu}(x^{a}_{1}, \ldots, y^{a}_{M})]
\nonumber\\
& &
A^{e}_{\nu}  + \bar{A}^{e}_{\nu}(x^{b}_{1}, \ldots, y^{b}_{M}) +
\bar{C}^{e}_{\nu}(x^{b}_{1}, \ldots, y^{b}_{M})]
[A^{f}_{\rho} + \bar{A}^{f}_{\rho}(x^{b}_{1}, \ldots, y^{b}_{M}) +
\bar{C}^{f}_{\rho}(x^{a}_{1}, \ldots, y^{a}_{M})]\Big]
+\mbox{Tr}D_{\mu}\Phi D^{\mu}\Phi
\nonumber\\
& &
- (4\lambda)^{2}\mbox{Tr}\Phi^{2}(\eta^{2} + \Phi^{2})^{2}
+{\cal L}_{GF}[A] + {\cal L}_{gh}[A]\Bigg]\Bigg\},
\nonumber
\end{eqnarray}
where ${\cal L}_{GF}$ and ${\cal L}_{gh}$ are respectively the gauge-fixing and ghost lagrangians.

We may obtain an equivalent expression for the $\sigma\mu$-correlation function, by shifting the
$A^{\mu}$ functional integration variable in the above equation as
\begin{eqnarray}
A^{\mu} \rightarrow A^{\mu} - \bar{A}^{\mu} - \bar{C}^{\mu}.
\end{eqnarray}

This produces the equivalent expression
\begin{eqnarray}
\label{fff}
& &
\langle \sigma(x^{a}_{1})\mu_{R}(x^{b}_{1})\ldots \sigma(x^{a}_{N})\mu_{R}(x^{b}_{N})\mu^{\dagger}_{R}(y^{b}_{M})\sigma^{\dagger}(y^{a}_{M})
\ldots\mu^{\dagger}_{R}(y^{b}_{1})\sigma^{\dagger}(y^{a}_{1})\rangle =
\nonumber\\
& &
{\cal Z}^{-1}\int {\cal D} A^{a}_{\mu}{\cal D}\Phi^{b}{\cal D}\eta {\cal D}
\bar{\eta}\,\exp\Bigg\{
-\frac{\kappa}{4\pi}\int d^{3}z\,\epsilon^{\mu\nu\rho}\left(
A^{a}_{\mu}\partial_{\nu}A^{a}_{\rho} +\frac{2}{3}\,\epsilon^{abc}A^{a}_{\mu}A^{b}_{\nu}
A^{c}_{\rho}\right)
+ \mbox{Tr}\bar{D}_{\mu}\Phi \bar{D}^{\mu}\Phi
\nonumber\\
& &
- (4\lambda)^{2}\mbox{Tr}\Phi^{2}(\eta^{2} +
\Phi^{2})^{2}
+{\cal L}_{GF}[A \rightarrow A^{\mu} - \bar{A}^{\mu} - \bar{C}^{\mu}]
+ {\cal L}_{gh}[A \rightarrow A^{\mu} - \bar{A}^{\mu} - \bar{C}^{\mu}]\Bigg\},
\end{eqnarray}
where
\begin{eqnarray}
\bar{D}_{\mu} = 1\partial_{\mu} + [A_{\mu} - \bar{A}_{\mu} - \bar{C}_{\mu}].
\end{eqnarray}

From this form of the correlation function we can extract the operators carrying, respectively, charge
and magnetic flux for the non-Abelian Chern-Simons-Higgs theory. This is easily done,
because, according to (\ref{fff}), a 2-point correlator of the $\mu$-operator, for instance, is expressed as a functional
integral having an integrand of the form:

\begin{eqnarray}
\label{mu-op}
 \exp\left\{-\int d^3x \left [ \frac{1}{2} W_\mu W^\mu 
+ W_\mu \bar{A}^\mu +\frac{1}{2} \bar{A}_\mu \bar{A}^\mu \right] \right\}
\end{eqnarray}
where $W_\mu$ is given in terms of the dynamical fields and $\bar{A}_\mu$ is given by
(\ref{campoexterno1}). The last term clearly does not involve dynamical fields, being
therefore a kind of renormalization factor. The first term is the measure weight, which is used for computing averages, namely

\begin{eqnarray}
\label{mu-av}
\langle\mu(x) \mu^\dagger (y)\rangle =
\int {\cal D} A^{a}_{\mu}{\cal D}\Phi^{b}{\cal D}\eta {\cal D}\bar{\eta}
 \exp\left\{-\int d^3x \left [ \frac{1}{2} W_\mu W^\mu 
 \right] \right\}
\mu(x) \mu^\dagger (y)
\end{eqnarray}

It follows that, appart from the (c-number) renormalization factor, the $\mu(x) \mu^\dagger(y)$ operators must
correspond to the second term in (\ref{mu-op}), which has an exponent linear in $\bar{A}_\mu (z;x,y)=
\bar{A}_\mu (z;x)-\bar{A}_\mu (z;y)$. The first one will give $\mu(x)$, whereas the second, $\mu^\dagger(y)$.

The $\sigma$-operators, conversely, must be exponentials of those terms that are linear in the external fields
$\bar{C}_\mu (z;x)$, given by (\ref{campoexterno2}).

Following the previously described inspection procedure, we obtain
\begin{eqnarray}
\label{mu-Higgs}
\mu(x_{i}) = \exp\left\{-n^{b} a\int_{x_{i}, L}^{+\infty} d\xi_{\mu}\epsilon^{\mu\alpha\nu}
\partial_{\nu}\,\frac{J_{\alpha}^{b}(\xi)}{(-\Box)}\right\}
\end{eqnarray}
and
\begin{eqnarray}
\sigma(x_{i}) = \exp\left\{n^{a}b\int_{x_{i}, L}^{+\infty} d\xi^{\mu}\,J^{a}_{\mu}(\xi)\right\}
\end{eqnarray}
with $J^{a}_{\mu}$ given by (\ref{EL}).

In the above equations, $ d\xi^{\mu}$ is the covariantized vector line integration element along
the integration line $L$. This is a  line going from $x_{i}$ to $\infty$ along the cut of the
$\mbox{arg}(z - x_{i})$ function.

We investigate now the commutation rules of $\sigma$ and $\mu$ operators obtained
above with the charge and magnetic flux operators.
Let us evaluate firstly the equal time commutator of $\mu$ with the magnetic flux operator.
Using the field equation, we may cast the $\mu$-operator in the form \cite{emijmp}
\begin{eqnarray}
\label{operator-mu}
\mu(x_{i}) = \exp\left\{\kappa a n^{b}\int_{x_{i}, L}^{+\infty}d\xi^{\mu}A^{b}_{\mu}(\xi)\right\}.
\end{eqnarray}
From this, we get
\begin{eqnarray}
[\mu(x_{i}), \Phi_{M}] &=&\frac{1}{2} \mu(x_{i})\kappa a\,n^{b}n^{c}\int d^{2}y\int_{x_{i}, L}^{+\infty} d\xi^k
[A^{b}_{k}(\xi), \epsilon^{jl}F_{jl}^{c}]
\nonumber\\
&=&- 2\pi a \mu(x_{i})\,n^{b}n^{c}\int d^{2}y\int_{x_{i}, L}^{+\infty} d\xi^k \partial_k^{(\xi)} \delta^{bc}
\delta(\xi - y)
\nonumber\\
& &
+\,\mu(x_{i})\kappa a\,n^{b}n^{c}\int d^{2}y\int_{x_{i}, L}^{+\infty} d\xi^k \epsilon^{jl}
\epsilon^{ced}[A^{b}_{k}(\xi), A^{e}_{j}(y)A^{d}_{l}(y)]
\nonumber\\
&=& 2\pi a\mu(x_{i}) + \mu(x_{i}) \kappa a n^{b}n^{c}\int d^{2}y\int_{x_{i}, L}^{+\infty}d\xi^k \epsilon^{jl}
\epsilon^{ced}\left(A^{e}_{j}(y)[A^{b}_{k}(\xi), A^{d}_{l}(y)] +
[A^{b}_{k}(\xi), A^{e}_{j}(y)]A^{d}_{l}(y)\right)
\nonumber\\
&=& 2\pi a\mu(x_{i})
\nonumber\\
& &
+ \mu(x_{i}) \kappa a n^{b}n^{c} \int d^{2}y\int_{x_{i}, L}^{+\infty} d\xi^k \epsilon^{jl}
\epsilon^{ced}\Big(A^{e}_{j}(y)\frac{2\pi}{\kappa}\epsilon_{kl}\delta^{bd}\delta^{2}(\xi - y)
+ \frac{2\pi}{\kappa}\epsilon_{kj}\delta^{be}\delta^{2}(\xi - y)A^{d}_{l}(y)\Big)
\nonumber\\
&=& 2\pi a\mu(x_{i})
\end{eqnarray}
where we used the equal-time commutator $[A^{a}_{i}(x), A^{b}_{j}(y)] = 2\pi/\kappa\,
\epsilon_{ij}\delta^{ab}\delta^{2}(\vec{x} - \vec{y})$ and the fact that
$n^{a}\delta^{ab} n^{b} = 1$. 
The second term in the rhs above vanishes because it is proportional to $n^{b}n^{c} \epsilon^{bcd}$.

This result shows that $\mu(x_{i})$ creates
states bearing a magnetic flux $2\pi a$, being therefore, a
magnetic vortex creation operator. Notice that $2\pi$ is the quantum of magnetic flux for $\hbar=c=e=1$, hence
the free parameter $a$ determines the number of flux units created by $\mu$. A natural choice, therefore would be
$a=1$.

In order to evaluate the commutator of $\sigma$ with the
matter charge operator $Q$, we must consider the current-current commutator. This is given in general by the current algebra relation
\cite{itzykson1980quantum}
\begin{eqnarray}
[J^{0 a}(\vec{x}, t), J^{i b}(\vec{y}, t)] = {\cal M}\delta^{ab}\partial^{i}\delta^{(2)}(
\vec{x} - \vec{y})
\end{eqnarray}
where ${\cal M}$ is a functional of the spectral density of the theory. Using this, we find
\begin{eqnarray}
[J^{0 a}(\vec{y}, t), \sigma(\vec{x}, t)] = b {\cal M}\sigma(\vec{x}, t)
\delta^{(2)}(\vec{x} - \vec{y})
\end{eqnarray}
or
\begin{eqnarray}
[Q, \sigma(\vec{x}, t)] = b {\cal M}\sigma(\vec{x}, t),
\end{eqnarray}
indicating that $\sigma$ bears a charge $b {\cal M}$
From this, we can see that the choice $b^{-1} = {\cal M}$ would imply that the operator
$\sigma$ carried one unit of electric charge.

We see that the $\mu$ and $\sigma$ operators obtained above carry, respectively, magnetic flux and charge.
We therefore expect their product will be, in general anyon fields \cite{fw}. It is precisely
out of combinations of these that we will construct the fields with non-Abelian statistics.

\subsection{Broken and Symmetric phases}
\subsubsection{Symmetric phase}
In the symmetric phase,  $\eta^{2} > 0$ in (\ref{CSH}), we have to add to (\ref{CSH}) a gauge-fixing term ${\cal L}_{GF}$
along with the corresponding ghost term ${\cal L}_{gh}$. For the gauge-fixing term, we are
going to choose a Lorentz-type gauge. Then, we add to (\ref{CSH}), in the symmetric phase, the
terms
\begin{eqnarray}
\label{gauge-symmetric}
{\cal L}^{S}_{GF} &=& -\frac{\xi}{2}\,(\partial_{\mu}A^{\mu a})^{2}
\nonumber\\
{\cal L}^{S}_{gh} &=& [\partial_{\mu}\bar{\eta}^{a}][D_{\mu}^{\tiny{adj}}\eta]^{a}
\end{eqnarray}
where $\eta^{a}$ are ghosts fields and $\xi$ is the gauge parameter.

\subsubsection{Broken phase}
In the broken phase, $\eta^{2} < 0$ in (\ref{CSH}), the potential has a minimum at
$\Phi^{2} = \phi^{a}_{0}$, with $\phi^{2}_{0} = \vert \eta^{2}\vert$. Taking the vacuum
pointing along the third direction, that is, $\phi^{a} = \phi_{0}\delta^{a3}$, we can see
that the fields will be given by ($\Phi^{1}, \Phi^{2}, \chi$), with $\chi = \Phi^{3} - \phi_{0}$. The
fields $\Phi_{1}$, $\Phi_{2}$, and $\chi$ have a zero vacuum expection value.

Then, in the broken phase we choose an 't Hooft gauge, where the quadratic mixed terms involving
($A_{\mu}, \Phi$) in the expression ${\cal L}^{B}$ disappear. To be more general, however, this unwanted
term disappears if we add a gauge-fixing term of the form
\begin{eqnarray}
\label{gauge-assymmetric}
{\cal L}^{B}_{GF} = -\frac{\xi}{2}\left[\partial_{\mu}A^{\mu a} +\frac{2M}{\xi}\,\epsilon^{ab3}\Phi_{b}
\right]^{2}.
\end{eqnarray}
where $M$ is the vacuum expectation value of the Higgs field.

From this gauge-fixing we have
\begin{eqnarray}
\label{ghosts-assymmetric}
{\cal L}^{B}_{gh} = [\partial_{\mu}\bar{\eta}^{a}][D_{\mu}^{\tiny{adj}}\eta]^{a} -
\bar{\eta}^{a}\left[\frac{2M}{\xi}\Phi_{3} + \frac{4M^{2}}{\xi}\right]
(\delta^{ab} - \delta^{a3}\delta^{b3})\eta^{b}.
\end{eqnarray}

From Eqs. (\ref{CSH}) and (\ref{gauge-symmetric}), we have the Langrangian density in the symmetric phase,
${\cal L}_{eff}^{S} = {\cal L}^{S} + {\cal L}_{GF}^{S} + {\cal L}_{gh}^{S}$, while in the broken phase, the Lagrangian density is ${\cal L}_{eff}^{B} =
{\cal L}^{B} + {\cal L}_{GF}^{B} + {\cal L}_{gh}^{B}$. From the quadratic terms in ${\cal L}_{eff}^{S}$ and
${\cal L}_{eff}^{B}$ we obtain the propagators for the fields. In Euclidean space these are
\begin{eqnarray}
\Delta_{(i)}(x) &=& \int\frac{d^{3}k}{(2\pi)^{3}}\frac{e^{ik\cdot x}}{k^{2} + m_{i}^{2}},\quad i = 1, 2, 3,
\nonumber\\
D^{\mu\nu}_{(1)}(x) &=& D^{\mu\nu}_{(2)}(x) = \int \frac{d^{3}k}{(2\pi)^{3}}
\frac{e^{ik\cdot x}}{2(\alpha^{2}k^{2} + M^{4})}\left[\alpha\epsilon^{\mu\lambda\nu}k_{\lambda}
+M^{2}\left(\delta^{\mu\nu} -
\frac{(\xi - 2\alpha^{2}/M^{2})k^{\mu}k^{\nu}}{(\xi k^{2} + 2M^{2})}\right)\right],
\nonumber\\
D^{\mu\nu}_{(3)}(x) &=& \int \frac{d^{3}k}{(2\pi)^{3}}
e^{ik\cdot x}\left[\frac{1}{2\alpha}\epsilon^{\mu\lambda\nu}\frac{k_{\lambda}}{k^{2}} +
\frac{1}{\xi}\frac{k^{\mu}k^{\nu}}{k^{4}}\right],
\nonumber\\
\Delta_{gh}^{(i)}(x) &=& \int\frac{d^{3}k}{(2\pi)^{3}}\frac{e^{ik\cdot x}}{k^{2} + m_{gh_{(i)}}^{2}},\quad i = 1, 2, 3,
\label{propagators}
\end{eqnarray}
where $\alpha = \kappa/4\pi$ and $\Delta_{(i)}$ are the propagators for the Higgs-field 
components, $\Phi_{1}$, $\Phi_{2}$, and
$\phi_{3}$ ($\chi$ in the broken phase), $D^{\mu\nu}_{(a)}(x)$ are the propagators for the gauge fields
$A^{a}_{\mu}$ and $\Delta^{(i)}_{gh}(x)$ are the propagators for ghosts-field components. In the
symmetric phase we have $M = 0$ and $m_{i}^{2} = (4\lambda)^{2}\vert \eta^{2}\vert^{2}$ and
$m^{2}_{gh_{(i)}} = 0$ ($i = 1, 2, 3$). In the broken phase we have $m_{1}^{2} = m_{2}^{2} = 4M^{2}/\xi$,
$m_{3}^{2} = m_{\chi}^{2}$, and $m_{gh_{(1)}}^{2} = m_{gh_{(2)}}^{2} = 4M^{2}/\xi$ and $m^{2}_{gh_{(3)}}= 0$.

\section{Vortex Correlation Functions}
\subsection{Introducing the external field in ${\cal L}_{eff} =
{\cal L} + {\cal L}_{GF} + {\cal L}_{gh}$}
Let us write the exponent in (\ref{fff}) as $S_{eff} = \int d^{3}z[{\cal L}^{Eucl}_{eff} +
\bar{{\cal L}}^{Eucl}_{eff}(\bar{A}_{\mu} + \bar{C}_{\mu})]$, where
$\bar{{\cal L}}^{Eucl}_{eff}(\bar{A}_{\mu} + \bar{C}_{\mu})$
contains all the dependence on the external field $\bar{A}_{\mu}(z; x_{1},\ldots, y_{M})$ and
$\bar{C}_{\mu}(z; x_{1},\ldots, y_{M})$.

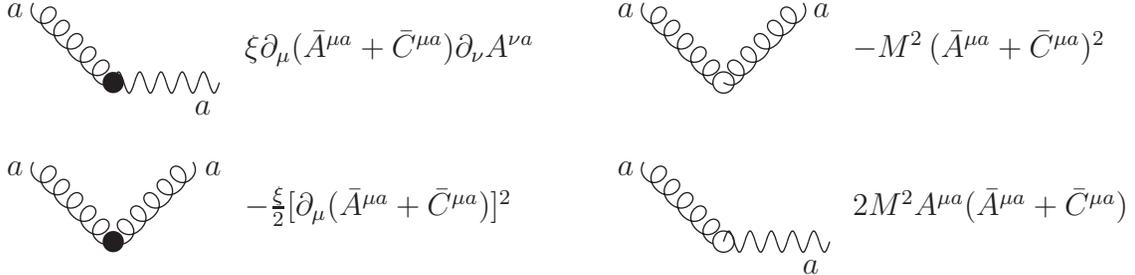
\begin{figure}[h!!]
\begin{center}
\begin{picture}(440,130)(0,0)
\Gluon(50,20)(20,50){4}{5}
\Vertex(50,20){4}
\Gluon(80,50)(50,20){4}{5}
\put(85,45){$a$}
\put(10,45){$a$}
\put(100,30){$-\frac{\xi}{2}[\partial_{\mu}(\bar{A}^{\mu a} + \bar{C}^{\mu a})]^{2}$}
\put(241,105){$a$}
\Gluon(280,80)(250,110){4}{5}
\BCirc(280,80){4}
\Gluon(310,110)(280,80){4}{5}
\put(315,105){$a$}
\put(330,90){$-M^{2}\,(\bar{A}^{\mu a} + \bar{C}^{\mu a})^{2}$}
\Gluon(50,80)(20,110){4}{5}
\Vertex(50,80){4}
\Photon(50,80)(90,80){4}{5}
\put(10,105){$a$}
\put(81,68){$a$}
\put(100,90){$\xi \partial_{\mu}(\bar{A}^{\mu a} + \bar{C}^{\mu a})\partial_{\nu}A^{\nu a}$}
\Gluon(280,20)(250,50){4}{5}
\BCirc(280,20){4}
\Photon(280,20)(320,20){4}{5}
\put(241,45){$a$}
\put(311,8){$a$}
\put(330,30){$2M^{2}A^{\mu a}(\bar{A}^{\mu a} + \bar{C}^{\mu a})$}
\end{picture}
{\caption{\small{Vertices involving the external field $\bar{A}^{\mu a} + \bar{C}^{\mu a}$ (curly line)
relevant for the evaluation of the mixed multi-correlation function. Those proportional to
$M^2$ only occur in the broken phase.}}
\label{Figura-1}}
\end{center}
\end{figure}

In the symmetric phase, from (\ref{fff}) and (\ref{gauge-symmetric}), we obtain that
\begin{eqnarray}
\hspace{-0.2cm}
\bar{{\cal L}}^{S}_{eff}(\bar{A}_{\mu} + \bar{C}_{\mu}) &=&
-2\epsilon^{abc}(\bar{A}_{\mu}^{b} + \bar{C}_{\mu}^{a})\Phi^{c}\partial_{\mu}\Phi^{a} +
(\bar{A}_{\mu}^{a} + \bar{C}_{\mu}^{a})(\bar{A}_{\mu}^{a} + \bar{C}_{\mu}^{a})\Phi^{b}\Phi^{b}
- 2(\bar{A}_{\mu}^{a} + \bar{C}_{\mu}^{a}A_{\mu}^{a})\Phi^{b}\Phi^{b}
\nonumber\\
& &
- (\bar{A}_{\mu}^{a} + \bar{C}_{\mu}^{a})(\bar{A}_{\mu}^{b} + \bar{C}_{\mu}^{b})\Phi^{a}\Phi^{b}
+ 2 (\bar{A}_{\mu}^{a} + \bar{C}_{\mu}^{a})A_{\mu}^{b}\Phi^{a}\Phi^{b} -\frac{\xi}{2}
[[\partial_{\mu}(\bar{A}_{\mu}^{a} + \bar{C}_{\mu}^{a})]^{2}
\nonumber\\
& & -
2\partial_{\mu}(\bar{A}_{\mu}^{a} + \bar{C}_{\mu}^{a})\partial_{\nu}(\bar{A}_{\nu}^{a} + \bar{C}_{\nu}^{a})]
-\bar{\eta}^{a}[\epsilon^{abc}\partial^{\mu}(\bar{A}_{\mu}^{c} + \bar{C}_{\mu}^{c}) +
\epsilon^{abc}(\bar{A}_{\mu}^{c} + \bar{C}_{\mu}^{c})\partial^{\mu}]\eta^{b}
\end{eqnarray}
while in the broken phase, we obtain
\begin{eqnarray}
\hspace{-0.8cm}
\bar{{\cal L}}^{B}_{eff}(\bar{A}_{\mu} + \bar{C}_{\mu}) &=&\frac{ M^{2}}{2}[(\bar{A}^{\mu}_{1} + \bar{C}^{\mu}_{1})^{2} -
2A^{\mu}_{1}(\bar{A}^{\mu}_{1} + \bar{C}^{\mu}_{1}) + (\bar{A}^{\mu}_{2} + \bar{C}^{\mu}_{2})^{2} -
2A^{\mu}_{2}(\bar{A}^{\mu}_{2} + \bar{C}^{\mu}_{2})]
\nonumber\\
& &
- [(\bar{A}^{\mu}_{1} + \bar{C}^{\mu}_{1})(\Phi_{2}\partial_{\mu}\chi - \chi\partial_{\mu}\Phi_{2})
+ (\bar{A}^{\mu}_{2} + \bar{C}^{\mu}_{2})(\chi\partial_{\mu}\Phi^{1} - \Phi^{1}\partial_{\mu}\chi)
\nonumber\\
& &
+ (\bar{A}^{\mu}_{3} + \bar{C}^{\mu}_{3})(\Phi^{1}\partial_{\mu}\Phi^{2} - \Phi^{2}\partial_{\mu}\Phi^{1})]
+
[(\bar{A}^{\mu}_{1} + \bar{C}^{\mu}_{1})^{2} - 2A^{\mu}_{1}(\bar{A}^{\mu}_{1} + \bar{C}^{\mu}_{1})]
(\Phi_{2}^{2}
\nonumber\\
& & + \chi^{2} + 2\phi_{0}\chi) + [(\bar{A}^{\mu}_{2} + \bar{C}^{\mu}_{2})^{2} - 2A^{\mu}_{2}(\bar{A}^{\mu}_{2} + \bar{C}^{\mu}_{2})](\Phi_{1}^{2} + \chi^{2} + 2\phi_{0}\chi)
\nonumber\\
& &
+[(\bar{A}^{\mu}_{3} + \bar{C}^{\mu}_{3})^{2} - 2A^{\mu}_{3}(\bar{A}^{\mu}_{3} + \bar{C}^{\mu}_{3})]
(\Phi^{2}_{1} + \Phi^{2}_{2}) - 2[(\bar{A}^{\mu}_{1} + \bar{C}^{\mu}_{1})(\bar{A}^{\mu}_{2} + \bar{C}^{\mu}_{2})
\nonumber\\
& &
A^{\mu}_{1}(\bar{A}^{\mu}_{2} + \bar{C}^{\mu}_{2}) - A^{\mu}_{2}(\bar{A}^{\mu}_{1} + \bar{C}^{\mu}_{1})]
\Phi_{1}\Phi_{2} - 2[(\bar{A}^{\mu}_{1} + \bar{C}^{\mu}_{1})(\bar{A}^{\mu}_{3} + \bar{C}^{\mu}_{3})
\nonumber\\
& &
A^{\mu}_{1}(\bar{A}^{\mu}_{3} + \bar{C}^{\mu}_{3}) - A^{\mu}_{3}(\bar{A}^{\mu}_{1} + \bar{C}^{\mu}_{1})]
(\Phi_{1}\chi + \phi_{0}\Phi_{1}) - 2[(\bar{A}^{\mu}_{2} + \bar{C}^{\mu}_{2})(\bar{A}^{\mu}_{3} + \bar{C}^{\mu}_{3})
\nonumber\\
& &
A^{\mu}_{2}(\bar{A}^{\mu}_{3} + \bar{C}^{\mu}_{3}) - A^{\mu}_{3}(\bar{A}^{\mu}_{2} + \bar{C}^{\mu}_{2})]
(\Phi_{2}\chi + \phi_{0}\Phi_{2})] - \frac{\xi}{2}
[[\partial_{\mu}(\bar{A}_{\mu}^{a} + \bar{C}_{\mu}^{a})]^{2}
\nonumber\\
& & -
2\partial_{\mu}(\bar{A}_{\mu}^{a} + \bar{C}_{\mu}^{a})\partial_{\nu}(\bar{A}_{\nu}^{a} + \bar{C}_{\nu}^{a})]
-\bar{\eta}^{a}[\epsilon^{abc}\partial^{\mu}(\bar{A}_{\mu}^{c} + \bar{C}_{\mu}^{c}) +
\epsilon^{abc}(\bar{A}_{\mu}^{c} + \bar{C}_{\mu}^{c})\partial^{\mu}]\eta^{b}.
\end{eqnarray}
From $\bar{{\cal L}}^{S}_{eff}(\bar{A}_{\mu} + \bar{C}_{\mu})$ and
$\bar{{\cal L}}^{B}_{eff}(\bar{A}_{\mu} + \bar{C}_{\mu})$, we can extract the Feynman rules
involving the external field. The relevant vertices are
shown in Fig. \ref{Figura-1}.

\subsection{The mixed correlation function}

The mixed correlation function can be expressed as
\begin{eqnarray}
\label{exponencial}
\langle \sigma(x^{a}_{1})\mu(x^{b}_{1})\ldots \sigma(x^{a}_{N})\mu(x^{b}_{N})\sigma^{\dagger}(y^{a}_{N})\mu^{\dagger}(y^{b}_{N})
\ldots\sigma^{\dagger}(y^{a}_{1})\mu^{\dagger}(y^{b}_{1})\rangle  = e^{-\Lambda(x^{a}_{1}, x^{b}_{1},\ldots, x^{a}_{N},x^{b}_{N};
y^{a}_{1}, y^{b}_{1},\ldots, y^{a}_{M},y^{b}_{M})},
\end{eqnarray}

It has been shown \cite{marino1992mass} that only the two legs graphs, containing the
external field $\bar{A}_{\mu} + \bar{C}_{\mu}$ will contribute to the large distance behavior of
$\Lambda(x^{a}_{1}, x^{b}_{1},\ldots, x^{a}_{N},x^{b}_{N};
y^{a}_{1}, y^{b}_{1},\ldots, y^{a}_{M},y^{b}_{M})$. 
At tree level the relevant graphs are depicted in Fig. \ref{figura-5}
In
the symmetric phase only the first two graphs in Fig \ref{figura-5} would contribute. Their sum, however,
actually vanishes, as we can see by using the gauge field propagators given in
Eq. (\ref{propagators}). This result leads us to the conclusion that in the symmetric phase we have
\begin{eqnarray}
\langle \sigma(x^{a}_{1})\mu_{R}(x^{b}_{1})\ldots \sigma(x^{a}_{N})\mu_{R}(x^{b}_{N})\mu^{\dagger}_{R}(y^{b}_{M})\sigma^{\dagger}(y^{a}_{M})
\ldots\mu^{\dagger}_{R}(y^{b}_{1})\sigma^{\dagger}(y^{a}_{1})\rangle_{S}
\stackrel{\vert \mathbf{x} - \mathbf{y}\vert \rightarrow \infty}{\sim} 1.
\end{eqnarray}
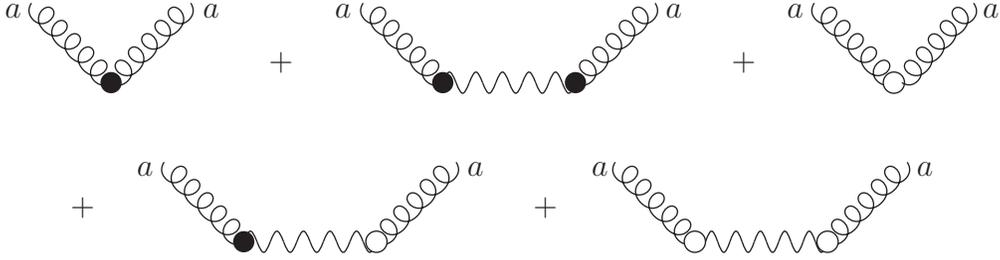
\begin{figure}[h!!]
\begin{center}
\begin{picture}(400,130)(0,0)
\put(10,105){$a$}
\Gluon(50,80)(20,110){4}{5}
\Vertex(50,80){4}
\Gluon(80,110)(50,80){4}{5}
\put(85,105){$a$}
\put(110,85){$+$}
\put(135,105){$a$}
\Gluon(175,80)(145,110){4}{5}
\Vertex(175,80){4}
\Photon(175,80)(225,80){4}{5}
\Gluon(255,110)(225,80){4}{5}
\Vertex(225,80){4}
\put(260,105){$a$}
\put(285,85){$+$}
\put(306,105){$a$}
\Gluon(345,80)(315,110){4}{5}
\BCirc(345,80){4}
\Gluon(375,110)(348,80){4}{5}
\put(380,105){$a$}
\put(35,30){$+$}
\put(60,45){$a$}
\Gluon(100,20)(70,50){4}{5}
\Vertex(100,20){4}
\Photon(100,20)(150,20){4}{5}
\Gluon(180,50)(150,20){4}{5}
\BCirc(150,20){4}
\put(185,45){$a$}
\put(210,30){$+$}
\put(232,45){$a$}
\Gluon(270,20)(240,50){4}{5}
\BCirc(270,20){4}
\Photon(274,20)(320,20){4}{5}
\Gluon(350,50)(320,20){4}{5}
\BCirc(320,20){4}
\put(355,45){$a$}
\end{picture}
{\caption{\small{Leading graphs contributing to the long distance behavior of
(\ref{exponencial}) in the non-Abelian CS-H theory. In the symmetric phase only
the first two appear.}}
\label{figura-5}}
\end{center}
\end{figure}

On the other hand, the last three graphs of Fig. \ref{figura-5} only occur in the broken
symmetry phase where the Higgs field possesses a nonzero vacuum expectation value, $M$.
From these, using the gauge propagators $D^{\mu\nu}_{de}(z)$  given in (\ref{propagators})
we can write explicitly

\begin{eqnarray}
& &
\Lambda(x^{a}_{1}, x^{b}_{1},\ldots, x^{a}_{N},x^{b}_{N};
y^{a}_{1}, y^{b}_{1},\ldots, y^{a}_{M},y^{b}_{M}) =
-M^{4}\sum_{a = 1}^{2}\int d^{3}zd^{3}z'\,
[\bar{A}^{d}_{\mu}(z; x^{b}_{1},\ldots, y^{b}_{M})
\label{26}
\\
& & \hspace{-0.8cm}
+ \bar{C}^{d}_{\mu}(z; x^{a}_{1},\ldots, y^{a}_{M})]\Big[i\alpha\epsilon^{
\mu\lambda\nu}\partial_{\lambda}
+\frac{\alpha^{2}}{M^{2}}\left(-\Box\delta^{\mu\nu} + \partial^{\mu}\partial^{\nu}\right)\Big]F(z - z')
[\bar{A}^{d}_{\mu}(z; x^{b}_{1},\ldots, y^{b}_{M}) + \bar{C}^{d}_{\mu}(z; x^{a}_{1},\ldots, y^{a}_{M})],
\nonumber
\end{eqnarray}
where
\begin{eqnarray}
F(z - z') = \int\frac{d^{3}k}{(2\pi)^{3}}\frac{e^{ik(z - z')}}{\alpha^{2}k^{2} + M^{4}}.
\end{eqnarray}

Expression (\ref{26}) has been evaluated in \cite{tese} giving

\begin{eqnarray}
&  &
\langle \sigma(x^{a}_{1})\mu_{R}(x^{b}_{1})\ldots \sigma(x^{a}_{N})\mu_{R}(x^{b}_{N})\mu^{\dagger}_{R}(y^{b}_{N})\sigma^{\dagger}(y^{a}_{N})
\ldots\mu^{\dagger}_{R}(y^{b}_{1})\sigma^{\dagger}(y^{a}_{1})\rangle  \stackrel{\vert \mathbf{x} - \mathbf{y}\vert \rightarrow \infty}{\sim}
\nonumber\\
& &
\exp\Bigg\{
- \pi a^{2}M^{2}\sum_{i, j= 1}^{N,N}\left(\vert x^{b}_{i} - y^{a}_{j}\vert -
\vert x^{b}_{i} - x^{a}_{j}\vert - \vert y^{b}_{i} - y^{a}_{j}\vert +
\vert y^{b}_{i} - x^{a}_{j}\vert\right)
\nonumber\\
& &
-4\pi iabM^{2}\sum_{i, j = 1}^{N,N}[\mbox{arg}({\bf x}^{b}_{i} -
{\bf y}^{a}_{j}) + \mbox{arg}({\bf y}^{b}_{i} - {\bf x}^{a}_{j})
- \mbox{arg}({\bf x}^{b}_{i} - {\bf x}^{a}_{j}) - \mbox{arg}({\bf y}^{b}_{i} - {\bf y}^{a}_{j})]\Bigg\}
\label{mcf}
\end{eqnarray}

Now, we can introduce a composite operator $\Psi(x)$ bearing charge and magnetic flux, through
\begin{eqnarray}
& &
\Psi(x) = \lim_{x^{a}, x^{b}\rightarrow x} \sigma(x^{a})\mu(x^{b})
\exp\Big\{-4\pi iabM^{2}\mbox{arg}({\bf x}^{b} - {\bf x}^{a})\Big\}.
\nonumber
\end{eqnarray}

From this and (\ref{mcf}) we obtain the large distance behavior of the composite operator correlation function:
\begin{eqnarray}
\label{phiphi}
&  &
\langle \Psi(x_{1})\ldots\Psi(x_{N})\Psi^{\dagger}(y_{N})\ldots\Psi^{\dagger}(y_{1})\rangle  \stackrel{\vert \mathbf{x} - \mathbf{y}\vert \rightarrow \infty}{\sim}
\exp\Bigg\{
-\pi a^{2}M^{2}\sum_{i, j= 1}^{N}\left(\vert x_{i} - y_{j}\vert +
\vert y_{i} - x_{j}\vert\right)
\nonumber\\
& &
+ \pi a^{2}M^{2}\sum_{i \neq j= 1}^{N}\left(
\vert x_{i} - x_{j}\vert + \vert y_{i} - y_{j}\vert\right)
-4\pi iabM^{2} \sum_{i, j = 1}^{N}[
\mbox{arg}({\bf x}_{i} - {\bf y}_{j}) + \mbox{arg}({\bf y}_{i} - {\bf x}_{j})]
\nonumber\\
& &      \hspace{6.5cm}
+ 4\pi iabM^{2} \sum_{i \neq j = 1}^{N}[\mbox{arg}({\bf x}_{i} - {\bf x}_{j}) +
\mbox{arg}({\bf y}_{i} - {\bf y}_{j})]\Bigg\}
\end{eqnarray}
where $\mbox{arg(z)} = \mbox{Arg(z)}+ 2\pi n $ and
we choose the cuts of the ${\mbox{Arg}}$ functions as
$-\pi \leqslant \mbox{Arg(z)} <  \pi$ and $0 \leqslant \mbox{Arg(-z)} < 2\pi$, in such a
way that we may write $ \mbox{Arg(- z)} = \mbox{Arg(z)}  + \pi$.

The composite field $\Psi$ carries magnetic flux and charge, which are both conserved quantities in the broken phase, hence the only non-vanishing functions
are the ones with the same number of operators and their Hermitean conjugates. This selection rule appears naturally in the
calculation leading to (\ref{phiphi}) \cite{tese,marinoap}.

The first term in (\ref{phiphi}) produces the exponential decay of the vortex correlation function.
This implies the energy spectrum of the quantum vortices $\Psi$ possesses a gap proportional
to the vacuum expectation value of the Higgs field squared: $M^2$. The fact that the above two-point function
vanishes asymptotically at large-distances means the quantum vortex states $|\Psi \rangle$ are orthogonal to the vacuum.
They are also orthogonal to isolated charge and magnetic flux states $|\sigma \rangle$ and $|\mu \rangle$.
These properties indicate that the $\Psi$-states are genuine and stable
quantum excitations of the system.

\subsection{Analytic Properties of the Correlation Functions}

The analytic structure of the eucllidean correlation functions is closely related to the spin 
value. Except for the case of bosons, the euclidean correlators are multivalued, each sheet 
corresponding to a different ordering of operators in the vacuum expectation values that 
correspond to these euclidean functions \cite{marswi}.

Observe that the composite field $\Psi$ (Euclidean) correlation function above is multivalued 
whenever $4\pi abM^{2} $ is not an integer. The composite field $\Psi$, indeed, has spin $s=4\pi abM^{2} $ 
and, as expected, is in general an anyon. Notice that $\Phi_M=2\pi a$  and $Q =b {\cal M}$ are 
respectively the magnetic flux  and the charge carried by the vortex operator. The spin, consequently, 
can be written in a more physical way as $s= 2\Phi_M QM^{2}  {\cal M}^{-1}$.

The values of the euclidean functions on adjacent sheets differ by a phase $e^{i 2 \pi s}$, hence, 
we have  the following property for the real time vacuum expectation values of fields \cite{marswi}.

\begin{eqnarray}
\label{braids}
\langle \Psi(x)\Psi^{\dagger}(y)\rangle^{(1)} = e^{i 2 \pi s} \langle \Psi^{\dagger}(y) \Psi(x)\rangle^{(1)} = e^{i 4 \pi s} \langle \Psi(x)\Psi^{\dagger}(y)\rangle^{(2)} = e^{i 6 \pi s}
\langle \Psi^{\dagger}(y) \Psi(x)\rangle^{(2)} = ...
\end{eqnarray}

Notice first that whenever $2 s = $ {\it integer} (bosons or fermions), $e^{i 2 \pi s} = \pm 1$. It follows 
that each of the vev's $\langle \Psi(x)\Psi^{\dagger}(y)\rangle $ and $\langle \Psi^{\dagger}(y) \Psi(x)\rangle$
is univalued in this case.

Notice now that, conversely, in the case of anyons, $2 s \neq$ {\it integer} and consequently the 
previous vev's are themselves multivalued, the particular sheet being indicated by the superscript. Observe 
that the values of the function in two adjacent sheets differ by a factor $e^{i 4 \pi s}$. This means
the vev's of field operators have a branch cut, the number of sheets being determined by the spin. For 
$ s = 1/N$  ($2 s \neq$ {\it integer}),  for instance, there are N sheets. For irrational spin, the 
vev's above would have an infinite number of sheets.

This analytic structure will be the basis for our costruction of states with Non-Abelian statistics.

\section{Fields with Non-Abelian Statistics}
\subsection{$2-$point correlation functions}
In this section, we show how to construct states with non-Abelian statistics out of the composite anyon vortex fields.
We start by considering the $2-$point correlation function of this field, namely
\begin{eqnarray}
\label{doispontos}
\langle \Psi(x)\Psi^{\dagger}(y)\rangle &=& \exp\left\{ -4\pi iabM^{2}[\mbox{Arg}(\bf{x} - \bf{y}) +
                                       \mbox{Arg}(\bf{y} - \bf{x})]\right\}\,e^{-D_{2}}
                                       \nonumber\\
                                       &=& e^{-2is\mbox{Arg}(\bf{x} - \bf{y})}e^{-is\pi}\,e^{-D_{2}}
\end{eqnarray}
in which $s = 4\pi abM^{2}$ and $D_{2} = 2\pi a^{2}M^{2}\vert x - y\vert $.

We now introduce the fields that will present non-Abelian statistics. For that purpose, let us 
consider the combined fields
\begin{eqnarray}
\label{29}
\Psi_{\pm}(x) = \frac{1}{2}(\Psi(x) \pm \Psi^{\dagger}(x)),
\end{eqnarray}
which are, respectively, self-adjoint and anti-self-adjoint. Using the fact that 
$\langle \Psi(x)\Psi(y)\rangle = \langle \Psi^{\dagger}(x)\Psi^{\dagger}(y)\rangle = 0$ we 
conclude that their correlation functions satisfy
\begin{eqnarray}
\langle\Psi_{+}(x)\Psi_{+}(y)\rangle =  -\langle
\Psi_{-}(x)\Psi_{-}(y)\rangle =
\frac{1}{4}\left(\langle \Psi(x)\Psi^{\dagger}(y)\rangle +
\langle \Psi^{\dagger}(x)\Psi(y)\rangle\right)
\end{eqnarray}
and
\begin{eqnarray}
\langle \Psi_{-}(x)\Psi_{+}(y)\rangle = - \langle
\Psi_{+}(x)\Psi_{-}(y)\rangle =
\frac{1}{4}\left(\langle \Psi(x)\Psi^{\dagger}(y)\rangle -
\langle \Psi^{\dagger}(x)\Psi(y)\rangle\right).
\end{eqnarray}

Using Eq. (\ref{doispontos}) we can write
\begin{eqnarray}
\label{mm}
\langle \Psi_{+}(x)\Psi_{+}(y)\rangle &=& \frac{1}{4}\left[e^{-2is\mbox{Arg}(\bf{x}
- \bf{y})}e^{-is\pi}e^{-D_{2}} +
e^{-2is\mbox{Arg}(\bf{y} - \bf{x})}e^{-is\pi}e^{-D_{2}}\right]
\nonumber\\
\langle \Psi_{-}(x)\Psi_{+}(y)\rangle &=& \frac{1}{4}\left[e^{-2is\mbox{Arg}(\bf{x}
- \bf{y})}e^{-is\pi}e^{-D_{2}} -
e^{-2is\mbox{Arg}(\bf{y} - \bf{x})}e^{-is\pi}e^{-D_{2}}\right]
\end{eqnarray}

Let us see what are the braiding properties of the states associated to the fields $\Psi_\pm$. Using
the properties of $\mbox{Arg (z)}$  in (\ref{mm})
we obtain
\begin{eqnarray}
& &
\left[e^{-2is\mbox{Arg}(\bf{x} - \bf{y})}e^{-is\pi}e^{-D_{2}} +
e^{-2is\mbox{Arg}(\bf{y} - \bf{x})}e^{-is\pi}e^{-D_{2}}\right]
\stackrel{\longrightarrow}{x \leftrightarrow y}
\Big[e^{-2\pi is}e^{-2is\mbox{Arg}(\bf{x} - \bf{y})}e^{-is\pi}e^{-D_{2}}
\nonumber\\
& & \hspace{10.0cm}
+\,\,e^{2\pi is}e^{-2is\mbox{Arg}(\bf{y} - \bf{x})}e^{-is\pi}e^{-D_{2}}\Big]
\nonumber
\end{eqnarray}
and
\begin{eqnarray}
& &
\left[e^{-2is\mbox{Arg}(\bf{x} - \bf{y})}e^{-is\pi}e^{-D_{2}} -
e^{-2is\mbox{Arg}(\bf{y} - \bf{x})}e^{-is\pi}e^{-D_{2}}\right]
\stackrel{\longrightarrow}{x \leftrightarrow y}
\Big[e^{-2\pi is}e^{-2is\mbox{Arg}(\bf{x} - \bf{y})}e^{-is\pi}e^{-D_{2}}
\nonumber\\
& & \hspace{10.0cm}
-\,\,e^{2\pi is}e^{-2is\mbox{Arg}(\bf{y} - \bf{x})}e^{-is\pi}e^{-D_{2}}\Big].
\nonumber
\end{eqnarray}

Observe that, whenever the operator $\Psi$ is bosonic or fermionic the phases generated by
braiding the $\Psi_{\pm}$-particles are identical, i. e., $e^{2\pi is} = e^{-2\pi is} = \pm 1$. This
implies
\begin{eqnarray}
\langle \Psi_{\pm}(y)\Psi_{\pm}(x)\rangle = e^{2\pi is}
\langle \Psi_{\pm}(x)\Psi_{\pm}(y)\rangle.
\nonumber
\end{eqnarray}
In this case, the above expression shows that whenever the charged vortex operator $\Psi$ is bosonic 
or fermionic, then the self-sdjoint operators $\Psi_{\pm}$ are also bosonic or fermionic.

On the other hand, when the vortex field is an anyon, namely, for $2s \neq $ {\it integer}, the $\Psi_{\pm}(x)$
fields have non-abelian braiding given by
\begin{eqnarray}
\langle \Psi_{+}(y)\Psi_{+}(x)\rangle &=& \frac{1}{4}\Big[\alpha^{*}\langle
(\Psi_{+}(x) + \Psi_{-}(x))(\Psi_{+}(y) - \Psi_{-}(y))\rangle
\nonumber\\
& & \quad
+\,\, \alpha\langle
(\Psi_{+}(x) - \Psi_{-}(x))(\Psi_{+}(y) + \Psi_{-}(y))\rangle\Big]
\nonumber\\
&=& \frac{1}{2}\Big[(\alpha + \alpha^{*})\langle \Psi_{+}(x)\Psi_{+}(y)\rangle -
(\alpha - \alpha^{*})\langle \Psi_{-}(x)\Psi_{-}(y)\rangle\Big]
\nonumber\\
&=& \cos \delta \langle \Psi_{+}(x)\Psi_{+}(y)\rangle - i\sin\delta\langle \Psi_{-}(x)\Psi_{-}(y)\rangle
\end{eqnarray}
and
\begin{eqnarray}
\langle \Psi_{-}(y)\Psi_{+}(x)\rangle &=& \frac{1}{4}\Big[\alpha^{*}\langle
(\Psi_{+}(x) + \Psi_{-}(x))(\Psi_{+}(y) - \Psi_{-}(y))\rangle
\nonumber\\
& & \quad \,\,
-\,\, \alpha\langle
(\Psi_{+}(x) - \Psi_{-}(x))(\Psi_{+}(y) + \Psi_{-}(y))\rangle\Big]
\nonumber\\
&=& \frac{1}{2}\Big[-(\alpha - \alpha^{*})\langle \Psi_{+}(x)\Psi_{+}(y)\rangle +
(\alpha + \alpha^{*})\langle \Psi_{-}(x)\Psi_{-}(y)\rangle\Big]
\nonumber\\
&=& -i\sin\delta\langle \Psi_{+}(x)\Psi_{+}(y)\rangle +
\cos\delta\langle \Psi_{-}(x)\Psi_{-}(y)\rangle
\end{eqnarray}
where in the above expression $\alpha = e^{i \delta}$ and $\delta = 2\pi s$.

We conclude that when the composite vortex field $\Psi$ is an anyon it follows that the $\Psi_{\pm}$ 
fields will have non-Abelian braiding given by
\begin{eqnarray}
\left( \begin{array}{c}
 \langle \Psi_{+}(y)\Psi_{+}(x)\rangle \\
 \langle \Psi_{-}(y)\Psi_{+}(x)\rangle  \\
\end{array} \right)
=
\left( \begin{array}{cc}
\cos\delta & -i\sin\delta \\
-i\sin\delta & \cos\delta \\
\end{array} \right)
\left(\begin{array}{c}
 \langle \Psi_{+}(x)\Psi_{+}(y)\rangle \\
 \langle \Psi_{-}(x)\Psi_{+}(y)\rangle  \\
\end{array} \right)
\nonumber
\end{eqnarray}

The braiding matrix and its hermitean adjoint
\begin{eqnarray}
\rho(M)=\left(\begin{array}{cc}
\cos\delta & -i\sin\delta \\
-i\sin\delta & \cos\delta \\
\end{array} \right)
\qquad
\rho(M)^{\dagger} = \left(\begin{array}{cc}
\cos\delta & i\sin\delta \\
i\sin\delta & \cos\delta \\
\end{array} \right)
\nonumber
\end{eqnarray}
satisfy $\rho(M)^{\dagger}\rho(M) = 1$, being therefore unitary.

We now come to one of our most interesting results:
Observe that a NOT gate can be obtained out of the braiding matrix $M$ (up to an i-factor)
by making $\delta = \pi/2$ or, equivalently, $s=1/4$, namely,
\begin{eqnarray}
M = -i X, \qquad \mbox{in which}\qquad X = \left( \begin{array}{cc}
0 & 1 \\
1 & 0 \\
\end{array} \right)
\end{eqnarray}

\subsection{$4$-point correlation functions}

Let us consider here the 4-point function of the vortex operator in the broken phase.
 From Eq. (\ref{phiphi}) we
can extract the following expression
\begin{eqnarray}
\langle \Psi(x_{1})\Psi(x_{2})\Psi^{\dagger}(x_{3})\Psi^{\dagger}(x_{4})\rangle &=&
\langle \Psi^{\dagger}(x_{1})\Psi^{\dagger}(x_{2})\Psi(x_{3})\Psi(x_{4})\rangle
\nonumber\\
&=&\exp\Big\{2is[\mbox{Arg}(\vec{x}_{1} - \vec{x}_{2}) - \mbox{Arg}(\vec{x}_{1} - \vec{x}_{3}) -
\mbox{Arg}(\vec{x}_{1} - \vec{x}_{4})
\nonumber\\
& &
- \mbox{Arg}(\vec{x}_{2} - \vec{x}_{3}) - \mbox{Arg}(\vec{x}_{2} - \vec{x}_{4}) +
\mbox{Arg}(\vec{x}_{3} - \vec{x}_{4})] - 2\pi is + C_{4a}\Big\}
\nonumber\\
\langle \Psi^{\dagger}(x_{1})\Psi(x_{2})\Psi^{\dagger}(x_{3})\Psi(x_{4})\rangle &=&
\langle \Psi(x_{1})\Psi^{\dagger}(x_{2})\Psi(x_{3})\Psi^{\dagger}(x_{4})\rangle
\nonumber\\
&=&\exp\Big\{2is[-\mbox{Arg}(\vec{x}_{1} - \vec{x}_{2}) + \mbox{Arg}(\vec{x}_{1} - \vec{x}_{3}) -
\mbox{Arg}(\vec{x}_{1} - \vec{x}_{4})
\nonumber\\
& &
- \mbox{Arg}(\vec{x}_{2} - \vec{x}_{3}) + \mbox{Arg}(\vec{x}_{2} - \vec{x}_{4}) -
\mbox{Arg}(\vec{x}_{3} - \vec{x}_{4})] - 2\pi is + C_{4b}\Big\}
\nonumber\\
\langle \Psi^{\dagger}(x_{1})\Psi(x_{2})\Psi(x_{3})\Psi^{\dagger}(x_{4})\rangle &=&
\langle \Psi(x_{1})\Psi^{\dagger}(x_{2})\Psi^{\dagger}(x_{3})\Psi(x_{4})\rangle
\nonumber\\
&=&\exp\Big\{2is[-\mbox{Arg}(\vec{x}_{1} - \vec{x}_{2}) - \mbox{Arg}(\vec{x}_{1} - \vec{x}_{3}) +
\mbox{Arg}(\vec{x}_{1} - \vec{x}_{4})
\nonumber\\
& &
+ \mbox{Arg}(\vec{x}_{2} - \vec{x}_{3}) - \mbox{Arg}(\vec{x}_{2} - \vec{x}_{4}) -
\mbox{Arg}(\vec{x}_{3} - \vec{x}_{4})] - 2\pi is + C_{4c}\Big\}
\nonumber\\
\label{1a}
\end{eqnarray}
where
\begin{eqnarray}
C_{4a} &=& -\pi a^{2}M^{2}\left(\vert x_{1} - x_{3}\vert + \vert x_{1} - x_{4}\vert
+ \vert x_{2} - x_{3}\vert + \vert x_{2} - x_{4}\vert \right)
\nonumber\\
& & +\,\pi a^{2}M^{2}\left(\vert x_{1} - x_{2}\vert + \vert x_{3} - x_{4}\vert\right)
\nonumber\\
C_{4b} &=& -\pi a^{2}M^{2}\left(\vert x_{4} - x_{3}\vert + \vert x_{4} - x_{1}\vert
+ \vert x_{2} - x_{3}\vert + \vert x_{2} - x_{1}\vert \right)
\nonumber\\
& & +\,\pi a^{2}M^{2}\left(\vert x_{4} - x_{2}\vert + \vert x_{3} - x_{1}\vert\right)
\nonumber\\
C_{4c} &=& -\pi a^{2}M^{2}\left(\vert x_{3} - x_{1}\vert + \vert x_{3} - x_{4}\vert
+ \vert x_{2} - x_{1}\vert + \vert x_{2} - x_{4}\vert \right)
\nonumber\\
& & +\,\pi a^{2}M^{2}\left(\vert x_{3} - x_{2}\vert + \vert x_{1} - x_{4}\vert\right).
\nonumber
\end{eqnarray}

The correlation functions of the new fields given by (\ref{29}) may be expressed in terms of the 
correlation functions above as
\begin{eqnarray}
\label{JKJ}
& & \hspace{-0.5cm}
\langle\Psi_{+}(x_{1})\Psi_{+}(x_{2})\Psi_{+}(x_{3})\Psi_{+}(x_{4})\rangle =
\langle \Psi_{-}(x_{1})\Psi_{-}(x_{2})\Psi_{-}(x_{3})\Psi_{-}(x_{4})\rangle =
\nonumber
\\
& & \hspace{-0.5cm}
2[\langle\Psi(x_{1})\Psi(x_{2})\Psi^{\dagger}(x_{3})\Psi^{\dagger}(x_{4})\rangle
+ \langle\Psi^{\dagger}(x_{1})\Psi(x_{2})\Psi^{\dagger}(x_{3})\Psi(x_{4})\rangle +
\langle\Psi^{\dagger}(x_{1})\Psi(x_{2})\Psi(x_{3})\Psi^{\dagger}(x_{4})\rangle]
\nonumber\\
\nonumber\\
& & \hspace{-0.5cm}
\langle \Psi_{+}(x_{1})\Psi_{+}(x_{2})\Psi_{-}(x_{3})\Psi_{-}(x_{4})\rangle =
\langle \Psi_{-}(x_{1})\Psi_{-}(x_{2})\Psi_{+}(x_{3})\Psi_{+}(x_{4})\rangle =
\nonumber\\
& & \hspace{-0.5cm}
2[\langle\Psi(x_{1})\Psi(x_{2})\Psi^{\dagger}(x_{3})\Psi^{\dagger}(x_{4})\rangle
- \langle\Psi^{\dagger}(x_{1})\Psi(x_{2})\Psi^{\dagger}(x_{3})\Psi(x_{4})\rangle -
\langle\Psi^{\dagger}(x_{1})\Psi(x_{2})\Psi(x_{3})\Psi^{\dagger}(x_{4})\rangle]
\nonumber\\
\nonumber\\
& &  \hspace{-0.5cm}
\langle \Psi_{-}(x_{1})\Psi_{+}(x_{2})\Psi_{-}(x_{3})\Psi_{+}(x_{4})\rangle =
\langle \Psi_{+}(x_{1})\Psi_{-}(x_{2})\Psi_{+}(x_{3})\Psi_{-}(x_{4})\rangle =
\nonumber\\
& &  \hspace{-0.5cm}
2[-\langle\Psi(x_{1})\Psi(x_{2})\Psi^{\dagger}(x_{3})\Psi^{\dagger}(x_{4})\rangle
+ \langle\Psi^{\dagger}(x_{1})\Psi(x_{2})\Psi^{\dagger}(x_{3})\Psi(x_{4})\rangle -
\langle\Psi^{\dagger}(x_{1})\Psi(x_{2})\Psi(x_{3})\Psi^{\dagger}(x_{4})\rangle]
\nonumber\\
\nonumber\\
& &   \hspace{-0.5cm}
\langle \Psi_{-}(x_{1})\Psi_{+}(x_{2})\Psi_{+}(x_{3})\Psi_{-}(x_{4})\rangle =
\langle \Psi_{+}(x_{1})\Psi_{-}(x_{2})\Psi_{-}(x_{3})\Psi_{+}(x_{4})\rangle =
\nonumber\\
& &   \hspace{-0.5cm}
2[-\langle\Psi(x_{1})\Psi(x_{2})\Psi^{\dagger}(x_{3})\Psi^{\dagger}(x_{4})\rangle
- \langle\Psi^{\dagger}(x_{1})\Psi(x_{2})\Psi^{\dagger}(x_{3})\Psi(x_{4})\rangle +
\langle\Psi^{\dagger}(x_{1})\Psi(x_{2})\Psi(x_{3})\Psi^{\dagger}(x_{4})\rangle]
\nonumber\\
\label{2a}
\end{eqnarray}

Let us see what are the braiding properties of the above functions. Using (\ref{1a}) and the 
expression above, we get

\begin{eqnarray}
& &
\langle \Psi_{+}(x_{1})\Psi_{+}(x_{2})\Psi_{+}(x_{3})\Psi_{+}(x_{4})\rangle
\stackrel{\longrightarrow}{x_{1} \leftrightarrow x_{2}} \quad
2\Big[e^{2\pi is}\langle \Psi(x_{1})\Psi(x_{2})\Psi^{\dagger}(x_{3})\Psi^{\dagger}(x_{4})\rangle
\nonumber\\
& & \hspace{1.0cm}
+\, e^{-2\pi is}\langle \Psi^{\dagger}(x_{1})\Psi(x_{2})\Psi(x_{3})\Psi^{\dagger}(x_{4})\rangle
+ e^{-2\pi is}\langle \Psi^{\dagger}(x_{1})\Psi(x_{2})\Psi^{\dagger}(x_{3})\Psi(x_{4})\rangle\Big]
\nonumber\\
\nonumber\\
& &
\langle \Psi_{+}(x_{1})\Psi_{+}(x_{2})\Psi_{-}(x_{3})\Psi_{-}(x_{4})\rangle
\stackrel{\longrightarrow}{x_{1} \leftrightarrow x_{2}} \quad
2\Big[ e^{2\pi is}\langle \Psi(x_{1})\Psi(x_{2})\Psi^{\dagger}(x_{3})\Psi^{\dagger}(x_{4})\rangle
\nonumber\\
& & \hspace{1.0cm}
- e^{-2\pi is}\langle \Psi^{\dagger}(x_{1})\Psi(x_{2})\Psi(x_{3})\Psi^{\dagger}(x_{4})\rangle
- e^{-2\pi is}\langle \Psi^{\dagger}(x_{1})\Psi(x_{2})\Psi^{\dagger}(x_{3})\Psi(x_{4})\rangle\Big]
\nonumber\\
\nonumber\\
& &
\langle \Psi_{-}(x_{1})\Psi_{+}(x_{2})\Psi_{-}(x_{3})\Psi_{+}(x_{4})\rangle
\stackrel{\longrightarrow}{x_{1} \leftrightarrow x_{2}} \quad
2\Big[- e^{2\pi is}\langle \Psi(x_{1})\Psi(x_{2})\Psi^{\dagger}(x_{3})\Psi^{\dagger}(x_{4})\rangle
\nonumber\\
& & \hspace{1.0cm}
+\, e^{-2\pi is}\langle \Psi^{\dagger}(x_{1})\Psi(x_{2})\Psi(x_{3})\Psi^{\dagger}(x_{4})\rangle
- e^{-2\pi is}\langle \Psi^{\dagger}(x_{1})\Psi(x_{2})\Psi^{\dagger}(x_{3})\Psi(x_{4})\rangle\Big]
\nonumber\\
\nonumber\\
& &
\langle \Psi_{-}(x_{1})\Psi_{+}(x_{2})\Psi_{+}(x_{3})\Psi_{-}(x_{4})\rangle
\stackrel{\longrightarrow}{x_{1} \leftrightarrow x_{2}} \quad
2\Big[- e^{2\pi is}\langle \Psi(x_{1})\Psi(x_{2})\Psi^{\dagger}(x_{3})\Psi^{\dagger}(x_{4})\rangle
\nonumber\\
& & \hspace{1.0cm}
- e^{-2\pi is}\langle \Psi^{\dagger}(x_{1})\Psi(x_{2})\Psi(x_{3})\Psi^{\dagger}(x_{4})\rangle
+ e^{-2\pi is}\langle \Psi^{\dagger}(x_{1})\Psi(x_{2})\Psi^{\dagger}(x_{3})\Psi(x_{4})\rangle\Big].
\label{E1}
\end{eqnarray}

Now with the help of the Eq. (\ref{2a}) we can write the right-hand side of Eq. (\ref{E1})
in terms of correlators of the new fields $\Psi_{+}$ and $\Psi_{-}$, namely
\begin{eqnarray}
\langle \Psi_{+}(x_{1})\Psi_{+}(x_{2})\Psi_{+}(x_{3})\Psi_{+}(x_{4})\rangle
\stackrel{\longrightarrow}{x_{1} \leftrightarrow x_{2}}
& &
\cos\delta\langle\Psi_{+}(x_{1})\Psi_{+}(x_{2})\Psi_{+}(x_{3})\Psi_{+}(x_{4})\rangle
\nonumber\\
& &
+\, i\sin\delta\langle \Psi_{+}(x_{1})\Psi_{+}(x_{2})\Psi_{-}(x_{3})\Psi_{-}(x_{4})\rangle
\nonumber\\
\langle \Psi_{+}(x_{1})\Psi_{+}(x_{2})\Psi_{-}(x_{3})\Psi_{-}(x_{4})\rangle
\stackrel{\longrightarrow}{x_{1} \leftrightarrow x_{2}}
& &
i\sin\delta\langle\Psi_{+}(x_{1})\Psi_{+}(x_{2})\Psi_{+}(x_{3})\Psi_{+}(x_{4})\rangle
\nonumber\\
& &
+\, \cos\delta\langle \Psi_{+}(x_{1})\Psi_{+}(x_{2})\Psi_{-}(x_{3})\Psi_{-}(x_{4})\rangle
\nonumber\\
\langle \Psi_{-}(x_{1})\Psi_{+}(x_{2})\Psi_{-}(x_{3})\Psi_{+}(x_{4})\rangle
\stackrel{\longrightarrow}{x_{1} \leftrightarrow x_{2}}
& &
 i\sin\delta\langle\Psi_{-}(x_{1})\Psi_{+}(x_{2})\Psi_{-}(x_{3})\Psi_{+}(x_{4})\rangle
\nonumber\\
& &
+\, \cos\delta\langle \Psi_{-}(x_{1})\Psi_{+}(x_{2})\Psi_{+}(x_{3})\Psi_{-}(x_{4})\rangle
\nonumber\\
\langle \Psi_{-}(x_{1})\Psi_{+}(x_{2})\Psi_{+}(x_{3})\Psi_{-}(x_{4})\rangle
\stackrel{\longrightarrow}{x_{1} \leftrightarrow x_{2}}
& &
 \cos\delta\langle\Psi_{-}(x_{1})\Psi_{+}(x_{2})\Psi_{-}(x_{3})\Psi_{+}(x_{4})\rangle
\nonumber\\
& &
+\, i\sin\delta\langle \Psi_{-}(x_{1})\Psi_{+}(x_{2})\Psi_{+}(x_{3})\Psi_{-}(x_{4})\rangle
\label{85}
\end{eqnarray}

From  (\ref{85}) we can determine the unitary matrix corresponding to
the braiding operation (monodromy matrix) $M_{12}$. Indeed, we may write the above equation as
\begin{eqnarray}
\left(\begin{array}{c}
\langle \Psi_{+}(x_{1})\Psi_{+}(x_{2})\Psi_{+}(x_{3})\Psi_{+}(x_{4})\rangle \\
\langle \Psi_{+}(x_{1})\Psi_{+}(x_{2})\Psi_{-}(x_{3})\Psi_{-}(x_{3})\rangle \\
\langle \Psi_{-}(x_{1})\Psi_{+}(x_{2})\Psi_{-}(x_{3})\Psi_{+}(x_{3})\rangle \\
\langle \Psi_{-}(x_{1})\Psi_{+}(x_{2})\Psi_{+}(x_{3})\Psi_{-}(x_{3})\rangle \\
\end{array} \right)
\stackrel{\longrightarrow}{x_{1} \leftrightarrow x_{2}}
\rho(\mbox{M}_{12})
\left(\begin{array}{c}
\langle \Psi_{+}(x_{1})\Psi_{+}(x_{2})\Psi_{+}(x_{3})\Psi_{+}(x_{4})\rangle \\
\langle \Psi_{+}(x_{1})\Psi_{+}(x_{2})\Psi_{-}(x_{3})\Psi_{-}(x_{3})\rangle \\
\langle \Psi_{-}(x_{1})\Psi_{+}(x_{2})\Psi_{-}(x_{3})\Psi_{+}(x_{3})\rangle \\
\langle \Psi_{-}(x_{1})\Psi_{+}(x_{2})\Psi_{+}(x_{3})\Psi_{-}(x_{3})\rangle \\
\end{array} \right)\nonumber
\end{eqnarray}
where
\begin{eqnarray}
\rho(\mbox{M}_{12}) =
\left(\begin{array}{cccc}
\cos\delta & i\sin\delta & 0 & 0 \\
i\sin\delta & \cos\delta & 0 & 0\\
0 & 0 & i\sin\delta & \cos\delta\\
0 & 0 & \cos\delta & i\sin\delta
\end{array} \right)\nonumber
\end{eqnarray}
and $\delta = 2\pi  s$. We see that it satisfies $\rho(\mbox{M}_{12})^{\dagger}\rho(\mbox{M}_{12}) = 1$, being
therefore unitary.

Using the same procedure we get the monodromy matrices that correspond to
the braiding operations $M_{13}$, $M_{14}$, $M_{23}$, $M_{24}$ and $M_{34}$. These are given by
\begin{eqnarray}
\rho(\mbox{M}_{34}) &=& \rho(\mbox{M}_{12}) \nonumber \\
\nonumber \\
\rho(\mbox{M}_{13}) &=& \rho(\mbox{M}_{24}) =
\left(\begin{array}{cccc}
\alpha^{*} & 0 & 0 & 0 \\
0 & 0 & 0 & \alpha^{*}\\
0 & 0 & \alpha^{*} & 0\\
0 & \alpha^{*} & 0 & 0
\end{array} \right)\nonumber
\end{eqnarray}
\begin{eqnarray}
\rho(\mbox{M}_{14}) =
\frac{1}{2}\left(\begin{array}{cccc}
\alpha^{*} + \beta^{*} & 0 & 0 & -\alpha^{*} + \beta^{*} \\
0 & -\alpha^{*} + \beta^{*} & \alpha^{*} + \beta^{*} & 0\\
0 & \alpha^{*} + \beta^{*} & -\alpha^{*} + \beta^{*} & 0\\
-\alpha^{*} + \beta^{*} & 0 & 0 & \alpha^{*} + \beta^{*}
\end{array} \right)\nonumber
\end{eqnarray}
{\small{\begin{eqnarray}
\rho(\mbox{M}_{23}) =
\left(\begin{array}{cccc}
\cos\delta & 0 & 0 & i\sin\delta \\
0 & i\sin\delta & \cos\delta & 0\\
0 & \cos\delta & i\sin\delta & 0\\
i\sin\delta & 0 & 0 & \cos\delta
\end{array} \right)
\quad
\nonumber
\end{eqnarray}}}
where $\alpha = e^{i\delta}$ and $\beta = e^{3i\delta}$.

The only commuting braiding matrices are $\rho(\mbox{M}_{14})$ and $\rho(\mbox{M}_{23})$, i. e.
\begin{eqnarray}
[\rho(\mbox{M}_{14}), \rho(\mbox{M}_{23})] = 0
\nonumber
\end{eqnarray}
It can be easily verified that the unitary monodromy braiding matrices satisfy the 
Yang-Baxter relations, 
\begin{eqnarray}
\rho(\mbox{M}_{12})\rho(\mbox{M}_{23})\rho(\mbox{M}_{12}) = 
\rho(\mbox{M}_{23})\rho(\mbox{M}_{12})\rho(\mbox{M}_{23}).
\end{eqnarray}
or, equivalently
\begin{eqnarray}
\rho(\mbox{M}_{23})\rho(\mbox{M}_{34})\rho(\mbox{M}_{23}) = 
\rho(\mbox{M}_{34})\rho(\mbox{M}_{23})\rho(\mbox{M}_{34}).
\end{eqnarray}

We now come to another of our most interesting results. Again we will see that for a particular choice of the spin,
the monodromy matrices become logic gates, which are essential for quantum computation
algorithms.
Indeed, a simply controlled-NOT operation (CNOT gate), then can be obtained by choosing $s = 1/4$ or
$\delta = \pi/2$ in $\rho(\mbox{M}_{12})$ (or $\rho(\mbox{M}_{34})$),  namely
\begin{eqnarray}
\rho(\mbox{M}_{12}) =
i\left(\begin{array}{cccc}
0 & 1 & 0 & 0 \\
1 & 0 & 0 & 0\\
0 & 0 & 1 & 0\\
0 & 0 & 0 & 1
\end{array} \right).
\nonumber
\end{eqnarray}

More logic keys may be obtained accordingly by a straightforward generalization.

At this point, one should inquire more precisely about what ultimately  determines the value of the spin of the vortices.
We have seen that $s= 2\Phi_M Q M^{2}  {\cal M}^{-1}$.  Assuming the vortices carry one unit of magnetic flux ($a=1$),
we have $\Phi_M= \frac{hc}{Q}$, hence the spin is

\begin{eqnarray}
s=2hc  M^{2}  {\cal M}^{-1},
\label{spin}
\end{eqnarray}
where we retrieved the physical units of magnetic flux. We conclude that the spin is determined by the ratio of the squared Higgs vacuum
expectation value to the current algebra scalar functional. The latter is a fixed number, determined by the spectral density of the theory.
 The former, is determined in principle by the Higgs potential parameters, however, in any
concrete associate condensed matter system, the Higgs expectation value is a physically
adjustable parameter, which will depend on the temperature of the system.
This means, therefore that the value of the spin s is ultimately determined by the temperature
and therefore could be adjusted to the value $1/4$ or to any other value by tuning the temperature appropriately.

\section{Conclusion}

We have shown that the 2+1 D Chern-Simons-Higgs theory in the broken phase  contains quantum states with non-Abelian statistics. These states
are created by operators that are combinations of electrically charged magnetic vortex fields and their Hermitean adjoints.
The Euclidean correlation functions of these composite
vortex operators have been obtained by the method of quantization of topological excitations, which is based on the idea of
order-disorder duality. All properties of such states may be derived from these correlation functions.
For instance, we may infer from their large distance behavior, that in the ordered phase
the quantum vortices are in excited quantum states carrying both
nonzero magnetic flux and charge and
having a gap proportional to the vacuum expectation value of the Higgs field.
We may also show, out of these correlation functions behavior, that the quantum vortices are in general anyons, as one should expect from
a field carrying both magnetic flux and charge.
 In special cases, however, they can be fermions or bosons, depending on the value of some input parameters.

The analytic properties of the Euclidean vortex correlation functions, by its turn, have been used in order to show that proper (self-adjoint or anti-self-adjoint) combinations of vortices and anti-vortices possess non-Abelian statistics, whenever these electrically charged vortices are anyons.
The unitary matrices corresponding to each of the non-commuting braiding
operations have been explicitly determined as a function of the spin $s$ for the case of the two and four-point correlation functions.

For a special value of the spin, namely $s=1/4$, we have shown that the monodromy matrices, which result from the exchange of the (anti) self-adjoint vortex states, become the basic logic gates (NOT, CNOT, and so on) required by the algorithm of a quantum computer. The spin value $s$, being proportional
to the vacuum expectation of the Higgs field can be tuned by the  temperature in a in any associated condensed matter system.

It would be nice to find a concrete material realization for the system studied here. Pure non-Abelian Chern-Simons theory has been claimed to describe the state corresponding to the plateau $\nu=5/2$ of Quantum Hall systems
\cite{qc-nas2,fradkin1998chern,moore1991nonabelions,naa1,naa3,naa4}.
It should be investigated, for that matter, whether the coupling of a Higgs field could have any physical meaning in this or in any related system.

\bigskip

\noindent{\bf Aknowledgments}

This work was supported in part by CNPq and FAPERJ

\end{document}